\documentclass[letterpaper, 10pt, conference]{IEEEtran}

\usepackage{subcaption}
\usepackage[utf8]{inputenc}
\usepackage[T1]{fontenc}
\usepackage{amsmath,amssymb}
\usepackage{graphicx}
\usepackage{booktabs}
\usepackage{siunitx}
\DeclareSIUnit\baud{Bd}
\DeclareSIUnit\dBm{dBm}
\usepackage{hyperref}
\usepackage{cite}
\usepackage{balance}
\usepackage{multirow}
\usepackage{xcolor}

\usepackage{graphics}

\csname title\endcsname{Towards Networked One Search Agent Systems: Multilateration of WiFi Fine Time Measurement Responders Using GNSS References}
\author{%
  \IEEEauthorblockN{Juan Bravo-Arrabal\textsuperscript{1},\quad
    Javier Ser\'on-Barba\textsuperscript{2},\quad
    Carlos~Sim\'on \'Alvarez-Merino\textsuperscript{3},}
  \IEEEauthorblockN{J.\,J.~Fern\'andez-Lozano\textsuperscript{2},\quad
    Alfonso Garc\'\i a-Cerezo\textsuperscript{2},\quad
      Anders Lyhne Christensen\textsuperscript{1}}
    
  \IEEEauthorblockA{\textsuperscript{1}SDU UAS Center, University of Southern Denmark, Odense, Denmark}
  \IEEEauthorblockA{\textsuperscript{2}Robotics and Mechatronics Lab, IMECH.UMA, University of M\'alaga, Spain}
  \IEEEauthorblockA{\textsuperscript{3}APMS Section, Aalborg University (AAU), Aalborg, Denmark}
  \IEEEauthorblockA{\texttt{juanba@mmmi.sdu.dk}}
}

\begin{document}
\maketitle

\begin{abstract}
This paper presents a proof-of-concept system for localising ground-based WiFi access points, acting as IEEE~802.11mc Fine Time Measurement (FTM) responders, from an uncrewed aerial vehicle using FTM ranging and Global Navigation Satellite System (GNSS)-referenced moving-baseline multilateration.
Each associated GNSS-referenced FTM-initiator pose supplies a known reference point, turning the flight trajectory into a temporal multilateration problem.
The real-time smartphone pipeline performs GNSS--ranging time association, robust outlier gating, a two-stage Gauss-Newton bootstrap, and sequential Bayesian filtering with bias tracking.
Six measurement-noise configurations, including empirical and adaptive models, are evaluated on field data collected in unstructured, mountainous terrain.
For a line-of-sight access point with \num{455} ranging measurements, the online Android pipeline achieves a final horizontal error of \SI{4.4}{\metre}, while offline replay of the same flight yields a time-weighted mean horizontal error of \SI{4.7}{\metre} and a best-case final horizontal error of \SI{1.1}{\metre} under the best noise model after a close flyby.
For non-line-of-sight targets, the real-time pipeline does not converge because of limited measurement availability, weak geometry, and signal attenuation, although an offline robust least-squares solver recovers a coarse solution for the vegetation-only case.
The system is intended as a building block for Networked One Search Agent architectures, and preliminary middleware tests demonstrate software-level interoperability, while quantitative multi-agent accuracy is left for future work.

\end{abstract}

\section{Introduction}\label{sec:intro}

\subsection{Motivation}

\begin{figure}[!h]
  \centering
  \IfFileExists{Figures/OSA-JEMERG.jpg}{%
    \includegraphics[width=\columnwidth]{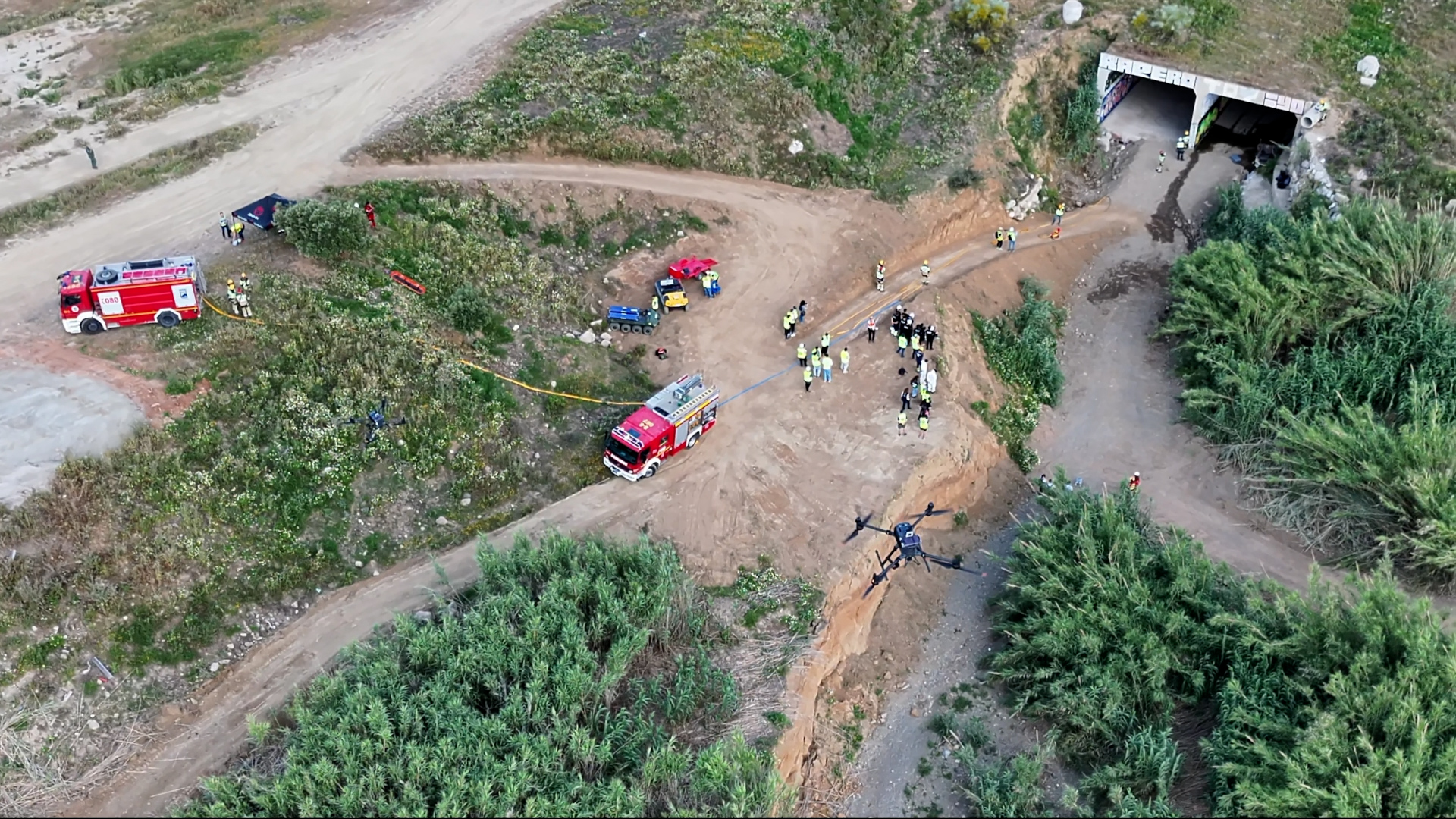}%
  }{%
    \fbox{\parbox[c][0.55\columnwidth][c]{0.9\columnwidth}{\centering Figures/OSA-JEMERG.jpg}}%
  }
    \caption{UAV operating as an OSA during XX JEMERG (May 8, 2026)~\cite{jemerg2026}, searching for victims carrying FTM responders.}
  \label{fig:osa_jemerg}
\end{figure}

Accurate localisation of stationary wireless emitters enables several robotic applications.
In search and rescue (SAR), victims and first responders carry WiFi-enabled devices whose positions can provide situational awareness during the initial \textit{Golden Hours}~\cite{lyu2023uav,queralta2020multi}.
Figure~\ref{fig:osa_jemerg} illustrates JEMERG~2026~\cite{jemerg2026}, where an uncrewed aerial vehicle (UAV) operating as a One Search Agent (OSA) searches for victims carrying \emph{Fine Time Measurement}~(FTM) responders.
In infrastructure mapping and persistent monitoring, WiFi access points (APs) can complement Global Navigation Satellite System (GNSS)-degraded navigation~\cite{pagliari2024wi}.
Across these scenarios, WiFi APs are already present or can be rapidly deployed, eliminating the need for dedicated positioning hardware.

Conventional approaches rely on signal-strength fingerprinting, which requires prior calibration, or dedicated ranging technologies such as Ultra-Wideband (UWB), which can achieve decimetre-level accuracy but demand purpose-built infrastructure~\cite{alarifi2016uwb,rykala2020research,queralta2020uwb}.
Robot-mounted radar has also been explored for emergency victim localisation~\cite{schroth2024emergency}, but similarly requires dedicated sensing equipment.
A middle ground is the IEEE~802.11mc FTM protocol: a \emph{Round-Trip Time}~(RTT) ranging method embedded in commodity WiFi hardware.
FTM yields metre-level range estimates without requiring clock synchronisation between initiator and responder, a property shared with UWB two-way ranging but achieved over standard WiFi channels~\cite{ftm_standard}.
Compared to UWB, WiFi FTM offers longer outdoor range (\SI{>100}{\metre} vs.\ \SIrange{30}{50}{\metre}) and uses existing infrastructure, but its narrower bandwidth limits time resolution and multipath resilience~\cite{alvarez2021wifi}.
The \SI{2.4}{\giga\hertz} ISM band spans only \SI{83.5}{\mega\hertz}, constraining devices to three non-overlapping \SI{20}{\mega\hertz} channels, whereas the \SI{5}{\giga\hertz} band offers roughly \SI{580}{\mega\hertz} and up to six non-overlapping \SI{80}{\mega\hertz} channels.

A newer WiFi positioning amendment, IEEE~802.11az, also known as Next Generation Positioning, extends FTM with improved ranging accuracy, wider bandwidth support, and enhanced protocol capabilities.
Android already provides platform support for WiFi RTT ranging through its WiFi RTT API, and Android~15 adds platform support for IEEE~802.11az-based ranging capabilities~\cite{android_rtt}.
In principle, 802.11az can provide higher ranging accuracy than 802.11mc, particularly under wideband configurations.

However, compatible responder hardware is not yet widely available for rapid field deployment, and current commodity access points predominantly support IEEE~802.11mc.
This work therefore focuses on IEEE~802.11mc-compatible devices, prioritising availability, simplicity, and deployability over best-case ranging precision.


\subsection{Related Work}

WiFi FTM has been extensively evaluated for indoor positioning~\cite{yu2020precise,ibrahim2018,ma2020wifi,martin2020passive}, with seamless indoor-outdoor extensions combining FTM and GNSS~\cite{yu2021seamless}.
Outdoor studies are more recent: Horn~\cite{horn2024outdoor} showed that RTT can outperform smartphone GNSS near buildings; Pagliari et~al.~\cite{pagliari2024wi} used FTM for UAV self-localisation with a fixed AP infrastructure; Lu et~al.~\cite{lu2024dnn} proposed deep-neural-network-based FTM corrections for UAVs; and Li et~al.~\cite{li2026coloc} addressed joint pedestrian-router colocalisation via factor graph optimisation.
In the multi-drone domain, Shaikhanov et~al.~\cite{shaikhanov2022falcon} demonstrated cooperative FTM-based target localisation.

Prior work by the authors progressively reduced infrastructure requirements for FTM-based victim localisation, from hybrid ZigBee/LoRa SAR networks~\cite{bravo2022realistic} and WiFi-based victim detection~\cite{alvarez2022victim,khatib2023designing} to FTM-based positioning.
In~\cite{bravo2025real}, a hybrid wireless sensor network of static and mobile anchors located victims at \SI{13}{\metre} median error.
The OSA concept~\cite{bravo2026osa} demonstrated that a single UAV can perform temporal multilateration with \SIrange{2}{6}{\metre} horizontal error.

\subsection{Contributions}

The original OSA concept~\cite{bravo2026osa} mounted an FTM anchor on the UAV and required each victim to carry a smartphone that computed RTT-based distances and forwarded them to a remote server for multilateration.
The present work advances that initial experimental prototype by inverting the localisation architecture: victims or first responders carry simple WiFi APs acting as FTM responders, while each OSA—implemented here as a commodity smartphone aboard the drone—acts as the FTM initiator, as well as GNSS receiver interface, on-board processor, and Internet gateway. This consolidation integrates ranging, positioning, estimation, and communication into a single device executing the multilateration algorithm in real time.
This inversion eliminates the need for software on the victim side, moves all computation to the search agent, and yields a lightweight, easy-to-deploy system capable of localising hidden or semi-hidden FTM transmitters in complex unstructured environments.
The objective is architectural rather than a same-dataset superiority claim over~\cite{bravo2026osa}: the present system obtains comparable metre-level line-of-sight performance while reducing the target-side requirement to an FTM-capable responder.
The main contributions are:
\begin{enumerate}
  \item A \textbf{complete real-time estimation pipeline} from raw FTM ranging to filtered 3-D AP position, including \emph{Median Absolute Deviation}~(MAD) outlier gating, a two-stage Gauss-Newton bootstrap, and sequential Bayesian filtering using an \emph{Iterated Extended Kalman Filter}~(IEKF) with bias tracking and an \emph{Unscented Kalman Filter}~(UKF), all running on a commodity smartphone.
  \item A \textbf{comparison of six measurement-noise configurations}, showing how distance- and \emph{Received Signal Strength Indicator}~(RSSI)-dependent weighting preserves measurement diversity and enables late close-pass corrections.
  \item \textbf{Proof-of-concept field validation} with \emph{line-of-sight}~(LoS) and \emph{non-line-of-sight}~(NLoS) access points under three graduated concealment conditions.
  \item \textbf{Preliminary software-level integration evidence} for a Networked OSA architecture, showing that heterogeneous UAV and ground agents can share FTM observations via Robot Operating System 2 (ROS\,2) in field tests, with field data and analysis scripts provided in a Zenodo reproducibility package; quantitative multi-agent positioning accuracy is not evaluated in this work.
\end{enumerate}

\section{System Architecture}\label{sec:system}

\subsection{Hardware and Software}

The UAV platform consists of a DJI Matrice 350 RTK airframe
carrying a consumer-grade Google Pixel 7 Pro smartphone.
The phone runs the \emph{RTT-OSA} Android application, whose source code will be shared on request.
The app interfaces simultaneously with the Android WiFi RTT API (IEEE~802.11mc) and a u-blox ZED-F9P multi-band GNSS receiver connected via USB-C at \SI{115200}{\baud}.
The F9P is configured at \SI{10}{\hertz} and supports \emph{Real-Time Kinematic}~(RTK) corrections via \emph{Networked Transport of RTCM via Internet Protocol}~(NTRIP) when available.
All sensor data is logged locally on the smartphone to structured newline-delimited JSON (JSONL) files, so the pipeline operates autonomously without network connectivity.
When a mobile data link is available, the app publishes observations through ROS\,2 Humble~\cite{ramos2024ur2a}, enabling remote monitoring and multi-agent fusion.
The smartphone is a pragmatic implementation choice rather than an architectural requirement; future work will compare companion-computer implementations and directional antennas.

The Android implementation maintains an independent track for each detected Basic Service Set Identifier (BSSID).
Every raw RTT result is logged and published through ROS\,2, but only measurements with a valid GNSS association and a GNSS quality proxy below \SI{15}{\metre} are passed to the estimator.
For each accepted BSSID track, the pipeline applies: (1)~monotonic-clock GNSS-RTT time association, (2)~optional pre-flight range-bias subtraction, (3)~MAD outlier gating, (4)~World Geodetic System 1984 (WGS84)-to-local-frame projection, (5)~two-stage Gauss-Newton bootstrap, (6)~sequential Bayesian filtering, (7)~soft ground-height regularisation, (8)~output smoothing, and (9)~JSONL/ROS\,2 publication.

\subsection{Range Measurement Model}

Let $\mathbf{p} = [x, y, z]^\top$ denote the unknown AP position in a local East-North-Up (ENU) frame, and let $\mathbf{u}_i=[x_i^d,y_i^d,z_i^d]^\top$ denote the smartphone/GNSS antenna position associated with the $i$-th FTM measurement.
Each measurement is therefore an end-to-end initiator--responder range from a known GNSS-referenced initiator pose to the unknown responder.
Thus, the known initiator poses form the virtual anchors for moving-baseline multilateration; in the previous OSA architecture the known endpoint was the UAV-mounted FTM responder, whereas here it is the UAV-mounted FTM initiator.
The local ENU frame is fixed at the first synchronised GNSS observation $(\phi_0,\lambda_0,h_0)$ using the same local tangent-plane approximation as the Android code:
\begin{equation}\label{eq:enu_projection}
\begin{split}
  x &= R_\oplus(\lambda-\lambda_0)\cos\!\left(\frac{\phi+\phi_0}{2}\right),\\
  y &= R_\oplus(\phi-\phi_0),\qquad z=h-h_0,
\end{split}
\end{equation}
where $\phi$, $\lambda$, and $h$ are the latitude, longitude, and height of the projected point, $\phi_0$, $\lambda_0$, and $h_0$ define the first synchronised GNSS observation, latitudes and longitudes are expressed in radians, and $R_\oplus=\SI{6378137}{\metre}$.
This approximation is sufficient for the few-hundred-metre spatial extent of the flights and avoids a heavier Earth-centred coordinate conversion on the smartphone.
For error evaluation and plotting, the RTK-surveyed AP ground-truth coordinates are projected into the same ENU frame, so estimated and reference positions are compared in a common local coordinate system.

The raw FTM range measurement is denoted by $d_i^{\mathrm{raw}}$, obtained from the RTT exchange between the initiator and the responder. The corrected range used by the estimator is $d_i=\max(0,d_i^{\mathrm{raw}}-b_0)$, where $b_0$ is an optional pre-flight calibration offset configured in the application.
The remaining observation model is:
\begin{equation}\label{eq:range_model}
  d_i = \|\mathbf{p} - \mathbf{u}_i\| + b + n_i,
\end{equation}
where $b$ is the residual constant FTM bias estimated online by the IEKF, and $n_i \sim \mathcal{N}(0, \sigma_i^2)$.
Successful responses arrive at \SIrange{3}{5}{\hertz} during active ranging.
FTM and GNSS are asynchronous; RTT timestamps are associated with GNSS fixes using Android's monotonic clock.
If bracketing GNSS fixes exist within $2T_\text{sync}$, the position is interpolated; otherwise, the closest fix is used only if it lies within $T_\text{sync}=\SI{1.2}{\second}$.
This timeout is inherited from the Android implementation rather than tuned as an estimator parameter: it sits above a \SI{1}{\hertz} fallback GNSS period, allowing modest Android callback jitter or one delayed fix while rejecting stale positions during longer outages.
With the ZED-F9P at \SI{10}{\hertz}, normal samples are bracketed much more tightly, so the \SI{1.2}{\second} limit mainly governs degraded or fallback association cases.
Unassociated RTT samples are logged but not used for multilateration.
The GNSS quality proxy $a_i$ is the application's \texttt{accM} field: for the external ZED-F9P stream, a Horizontal Dilution of Precision (HDOP)-derived horizontal accuracy proxy rather than a calibrated covariance.
It is used for gating and diagnostics, while Table~\ref{tab:results_1f} omits a separate GNSS covariance term so that the six FTM noise configurations remain comparable.

\subsection{MAD Outlier Gating}

A sliding-window MAD filter rejects gross outliers~\cite{huber2009}. Let $\tilde d$ be the median of the recent range window and $\mathrm{MAD} = \mathrm{median}_j(|d_j - \tilde d|)$ its median absolute deviation.
The window spans \SI{1500}{\milli\second} with $N_\text{min} = 8$ warm-up samples.
Once $\ge 8$ entries are available, the gate rejects sample~$d_i$ if:
\begin{equation}\label{eq:mad}
  |d_i - \tilde{d}| > k \cdot 1.4826 \cdot \mathrm{MAD},
\end{equation}
where the factor $1.4826 = 1/\Phi^{-1}(3/4)$ makes the MAD a consistent estimator of the standard deviation under Gaussian noise~\cite{huber2009}.
The threshold $k=3.5$ follows Iglewicz and Hoaglin~\cite{iglewicz1993} and preserves valid LoS measurements while rejecting spurious RTT outliers.
Accepted samples use the robust scale $\hat{\sigma}_\text{MAD} = 1.4826 \cdot \mathrm{MAD}$:
\begin{equation}\label{eq:sigma_gate}
  \sigma_\text{gate} = \max\!\bigl(\hat{\sigma}_\text{MAD},\; \sigma_\text{floor}\bigr),
\end{equation}
with $\sigma_\text{floor} = \SI{0.12}{\metre}$, which prevents variance collapse when consecutive measurements are unusually consistent.
Because the drone is moving, the window dispersion contains both FTM noise and genuine geometric range variation; $\sigma_\text{gate}$ is therefore a robust local scale for gating and weighting, not a pure sensor-noise estimate.
The default measurement variance is $R_k = \sigma_\text{gate}^2$.

\subsection{Gauss-Newton Bootstrap}\label{sec:bootstrap}

Once $\ge 8$ gated samples with geometric spread $\ge \SI{15}{\metre}$ (bounding-box diagonal of the drone positions in the horizontal plane) have accumulated, a two-stage solver computes the initial AP estimate.
This bootstrap provides a local initialisation for the subsequent IEKF or UKF; a poor initialisation can cause later innovation gates to reject otherwise valid measurements.

\textbf{Stage~1: Linear 2-D initialisation.}
The slant RTT ranges are first converted into approximate horizontal ranges using the configured flight height \emph{above ground level}~(AGL):
\begin{equation}\label{eq:agl_correction}
  \rho_i^2 = d_i^2 - H_\text{AGL}^2,
\end{equation}
where $H_\text{AGL}$ is the estimated altitude difference between the drone and the ground.
If the right-hand side becomes non-positive because of noise or an overestimated AGL value, the implementation falls back to $\rho_i^2=d_i^2$.
Subtracting the squared range equation of the first accepted sample from sample $i$ gives a linear equation in the horizontal AP coordinates:
\begin{multline}\label{eq:linear_2d}
  2(x_i^d-x_1^d)\,x + 2(y_i^d-y_1^d)\,y \\
  = \rho_1^2-\rho_i^2
    + (x_i^d)^2-(x_1^d)^2
    + (y_i^d)^2-(y_1^d)^2.
\end{multline}
Stacking these equations for $i=2,\ldots,N$, where $N$ is the number of gated samples used in the bootstrap, yields $\mathbf{A}\mathbf{p}_{xy}=\mathbf{c}$, solved by weighted least squares as
\begin{equation}\label{eq:linear_wls}
  \hat{\mathbf{p}}_{xy}=(\mathbf{A}^\top\mathbf{W}\mathbf{A})^{-1}\mathbf{A}^\top\mathbf{W}\mathbf{c},
\end{equation}
with $\mathbf{W}=\mathrm{diag}(1/\sigma_i^2)$, where $\sigma_i$ is the per-sample range-noise scale provided by the MAD gate.
The initial vertical coordinate is set to the robust ground-height estimate:
\begin{equation}\label{eq:z_initial}
  \hat z_0 = \mathrm{median}_i\,(z_i^d-H_\text{AGL}).
\end{equation}

\textbf{Stage~2: 3-D Gauss-Newton refinement.}
Starting from $[\hat{\mathbf{p}}_{xy}^\top,\hat z_0]^\top$, the solver minimises the nonlinear residuals:
\begin{equation}\label{eq:gn_residual}
  e_i(\mathbf{p}) = \|\mathbf{p}-\mathbf{u}_i\|-d_i.
\end{equation}

At each iteration the update is:
\begin{equation}\label{eq:gn_update}
  \Delta\mathbf{p} = -\left(\mathbf{J}^\top \mathbf{W} \mathbf{J}\right)^{-1} \mathbf{J}^\top \mathbf{W}\, \mathbf{e},
\end{equation}
where $J_{ij}=(p_j-u_{i,j})/\|\mathbf{p}-\mathbf{u}_i\|$ for coordinate index $j\in\{x,y,z\}$.
The estimate is then updated as:
\begin{equation}\label{eq:gn_state_update}
  \mathbf{p}^{(m+1)} = \mathbf{p}^{(m)} + \Delta\mathbf{p}^{(m)},
\end{equation}
where $m$ is the Gauss-Newton iteration index.
Convergence is declared when $\|\Delta\mathbf{p}\| < 10^{-4}$\,m or after 10~iterations, and solutions farther than $2\times$ the maximum observed range from the drone-position centroid are rejected as divergent.

\subsection{Sequential Bayesian Filtering}

The bootstrap estimate initialises one of two sequential filters, selected by an experiment configuration parameter set before each flight.
The IEKF is used in five of six modes; the UKF serves as a bias-free alternative for comparison.

\textbf{IEKF3D.}
An IEKF with state $\mathbf{x}_k = [x, y, z, b]^\top$ estimates both the AP position and a residual range bias.
The AP is assumed static, so prediction only increases the covariance:
\begin{equation}\label{eq:iekf_prediction}
  \mathbf{x}_k^- = \mathbf{x}_{k-1}^+,\qquad
  \mathbf{P}_k^- = \mathbf{P}_{k-1}^+ + \mathbf{Q},
\end{equation}
with $\mathbf{Q} = \mathrm{diag}(0.01,\, 0.01,\, 0.04,\, 10^{-4})\;\si{\metre^2}$.
The initial covariance after the bootstrap is $\mathbf{P}_0=\mathrm{diag}(25,25,100,4)$, corresponding to a deliberately broad prior of \SI{5}{\metre} in horizontal position, \SI{10}{\metre} in height, and \SI{2}{\metre} in residual bias.
The range measurement function is:
\begin{equation}\label{eq:meas_model}
  h(\mathbf{x}_k) = \sqrt{(x - x_k^d)^2 + (y - y_k^d)^2 + (z - z_k^d)^2} + b.
\end{equation}

The associated Jacobian matrix is:
\begin{equation}\label{eq:iekf_jacobian}
  \mathbf{H}_k = \left[
  \frac{x-x_k^d}{\hat d_k},\;
  \frac{y-y_k^d}{\hat d_k},\;
  \frac{z-z_k^d}{\hat d_k},\;1
  \right],
\end{equation}
where $\hat d_k=\|[x,y,z]^\top-\mathbf{u}_k\|$ is the predicted geometric range before adding the bias term.
The Jacobian is recomputed at each of $M = 3$ IEKF inner iterations, which improves consistency when the current estimate is not yet close to the AP.
Each incoming measurement is screened by a \emph{Normalised Innovation Squared}~(NIS) gate before the state is updated.
The scalar innovation is $\nu_k = d_k - h(\mathbf{x}_k^{-})$, observed minus predicted range, and its variance is:
\begin{equation}\label{eq:innov_cov}
  S_k = \mathbf{H}_k \mathbf{P}_k^{-} \mathbf{H}_k^\top + R_k,
\end{equation}
where $\mathbf{P}_k^{-}$ is the predicted state covariance before incorporation of measurement~$k$.
The NIS statistic $\gamma_k = \nu_k^2 / S_k$ follows a $\chi^2$ distribution with one degree of freedom under the filter's Gaussian assumptions; the gate rejects the update when
\begin{equation}\label{eq:nis}
  \gamma_k > 9.0,
\end{equation}
i.e.\ a three-standard-deviation scalar gate ($P(\chi^2_1>9)\approx0.0027$).
Accepted updates use the standard Kalman gain and the Joseph covariance form, where $\mathbf{I}$ denotes the identity matrix:
\begin{equation}\label{eq:joseph_update}
\begin{split}
  \mathbf{K}_k &= \mathbf{P}_k^-\mathbf{H}_k^\top S_k^{-1},\\
  \mathbf{x}_k^+ &= \mathbf{x}_k^- + \mathbf{K}_k\nu_k,\\
  \mathbf{P}_k^+ &= (\mathbf{I}-\mathbf{K}_k\mathbf{H}_k)\mathbf{P}_k^-(\mathbf{I}-\mathbf{K}_k\mathbf{H}_k)^\top
  + \mathbf{K}_k R_k \mathbf{K}_k^\top.
\end{split}
\end{equation}

The common subtractive update, $\mathbf{P}_k^+=(\mathbf{I}-\mathbf{K}_k\mathbf{H}_k)\mathbf{P}_k^-$, is obtained from the Joseph expression only after expanding it and cancelling terms using $S_k=\mathbf{H}_k\mathbf{P}_k^-\mathbf{H}_k^{\mathrm{T}}+R_k$ and the exact-gain identity $\mathbf{K}_kS_k=\mathbf{P}_k^-\mathbf{H}_k^{\mathrm{T}}$.
Under finite precision, these cancellations are not exact: repeated scalar range updates may leave a slightly asymmetric covariance or introduce small negative eigenvalues after subtracting nearly equal terms.
Therefore, the Joseph form avoids this fragile cancellation and keeps the update as the sum of positive semi-definite terms when $\mathbf{P}_k^-$ and $R_k$ are positive semi-definite, making invalid covariance matrices less likely; for this four-state IEKF, the extra cost is negligible.

After each accepted range update, both IEKF and UKF receive a soft pseudo-measurement of the ground height under the drone:
\begin{equation}\label{eq:z_prior}
  z_k^{\mathrm{prior}}=\mathrm{median}_{j\in\mathcal{W}_z}(z_j^d-H_\text{AGL}),
  \qquad R_z = (\SI{5}{\metre})^2.
\end{equation}

The set $\mathcal{W}_z$ denotes a sliding window of recent drone altitude samples used to estimate the ground height.
This prior does not use the surveyed AP height or force the state to ground truth; it centres the vertical estimate around $z_j^d-H_\text{AGL}$ with a \SI{5}{\metre} standard deviation, limiting drift of the weakly observable vertical state under near-constant-altitude flight.

\textbf{UKF3D.}
A three-state UKF ($[x, y, z]^\top$, no bias) with Merwe-scaled sigma points ($\alpha=10^{-3}$, $\beta=2$, $\kappa=0$).
It uses the same static process-noise values for position as the IEKF and the same NIS gate, but propagates $2n+1=7$ sigma points through the nonlinear range function, with state dimension $n=3$, instead of relying on an explicit Jacobian.
Because no bias state is present, the UKF serves as a comparison of nonlinear propagation against bias tracking rather than as a full replacement for the IEKF.

For both filters, the uncertainty indicators reported by the application are extracted from the current filter covariance before output smoothing:
\begin{equation}\label{eq:sigma_h_z}
  \sigma_H = \sqrt{P_{xx}+P_{yy}},\qquad
  \sigma_Z = \sqrt{P_{zz}}.
\end{equation}
Thus, $\sigma_H$ and $\sigma_Z$ describe internal filter uncertainty, whereas the horizontal and 3-D errors in Table~\ref{tab:results_1f} are computed by comparing the published position estimate with the RTK-surveyed AP ground truth.

\subsection{Measurement-Noise Configurations}

Six configurations vary how the FTM component of the measurement variance, $R_k^{\mathrm{FTM}}$, is computed.
In the offline replay used for the results reported in this paper, the filter update uses the mode-specific FTM variance without adding a separate GNSS covariance term.
The six modes are:
\begin{enumerate}
  \item \textsc{Baseline}: $R_k^{\mathrm{FTM}} = \sigma_\text{gate}^2$ (constant from MAD).
  \item \textsc{NLoS}: inflates the measurement noise standard deviation by $\kappa_\text{NLoS} = 3$ (equivalently, $R_k^{\mathrm{FTM}} \leftarrow 9\,R_k^{\mathrm{FTM}}$) when NLoS is detected.
  The detector combines range-window skewness and RSSI consistency. The skewness test is:
    \begin{equation}\label{eq:skewness}
      \gamma_1 = \frac{1}{N}\sum_{j=1}^{N}\left(\frac{d_j-\bar d}{s_d}\right)^3
    \end{equation}
  with threshold $\gamma_\mathrm{th}=1.0$ and $N\leq20$. The RSSI reference is:
    \begin{equation}\label{eq:rssi_path_loss}
      \mathrm{RSSI}_\text{exp}(d)=\mathrm{RSSI}_{1\mathrm{m}}-10\eta\log_{10}(d),
    \end{equation}
  with $\mathrm{RSSI}_{1\mathrm{m}}=\SI{-35}{\dBm}$, $\eta=3$, and corrected FTM range $d_k$. The combined decision and variance inflation are:
    \begin{equation}\label{eq:nlos_decision}
    \begin{split}
      I_\mathrm{NLoS} &= \mathbb{I}\!\left[\gamma_1>1.0\;\lor\;
      \left|\mathrm{RSSI}_k-\mathrm{RSSI}_\text{exp}(d_k)\right|>\SI{15}{\decibel}\right],\\
      R_k^{\mathrm{FTM}} &= \left(1+8I_\mathrm{NLoS}\right)\sigma_\text{gate}^2.
    \end{split}
    \end{equation}

  \item \textsc{Empirical-R}: distance- and RSSI-dependent noise,
    \begin{equation}\label{eq:empirical_R}
    \begin{split}
      R_k^{\mathrm{FTM}} &= \max\!\left(\sigma_\text{gate}^2,\;
      \sigma_0^2 + \alpha\, d_k^2 + \beta\, q_k^2\right),\\
      q_k &= \max\!\bigl(0,\, -\mathrm{RSSI}_k-r_\text{ref}\bigr),
    \end{split}
    \end{equation}
    with $\sigma_0=\SI{0.3}{\metre}$, $\alpha=4\times10^{-4}$, $\beta=3\times10^{-3}$, and $r_\text{ref}=\SI{40}{\decibel}$.
    The terms model the FTM floor~\cite{ibrahim2018}, distance-dependent multipath~\cite{horn2024outdoor}, and RSSI attenuation; the outer maximum prevents assigning less variance than the MAD gate.
    Coefficients were frozen before the field experiments, and no parameter was adjusted on the evaluation dataset.
  \item \textsc{UKF}: UKF3D with constant $R_k^{\mathrm{FTM}}$.
  \item \textsc{EMA}: IEKF with \emph{Exponential Moving Average}~(EMA) smoothing ($\alpha_\text{EMA} = 0.3$).
  \item \textsc{Adaptive-R}: Mohamed-Schwarz~\cite{mohamed1999} innovation-based adaptation, where the running measurement-variance estimate $\hat R_k$ is updated from the innovation covariance with explicit clipping:
    \begin{multline}\label{eq:adaptive_R}
      \hat{R}_k = \mathrm{clip}_{[0.01,\,25.0]}\!\bigl(
        (1 - \alpha_a)\,\hat{R}_{k-1} \\
        + \alpha_a\bigl(\nu_k^2 - \mathbf{H}_k \mathbf{P}_k^{-} \mathbf{H}_k^\top\bigr)
      \bigr),
    \end{multline}
    with adaptation rate $\alpha_a = 0.05$. Clipping enforces $\hat{R}_k \in [0.01,\,25.0]\;\si{\metre^2}$ ($\sigma \in [\SI{0.1}{\metre},\,\SI{5.0}{\metre}]$) to prevent noise collapse or divergence, and the clipped $\hat R_k$ is used as $R_k$.
\end{enumerate}

All modes except \textsc{EMA} use a sliding-window median smoother on the published filter output ($W_{xy}=7$ for horizontal coordinates and $W_z=9$ for height); the ground-height pseudo-measurement in~\eqref{eq:z_prior} uses a separate $25$-sample median.
Table~\ref{tab:results_1f} mixes aggregate counts, a final-sample snapshot, and one temporal summary.
The columns $n_\text{gate}$ and $n_\text{acc}$ are post-bootstrap counts over the whole replay; $\sigma_H$, $\sigma_Z$, $\varepsilon_\text{2D}$, and $\varepsilon_\text{3D}$ are the values at the final reported stable estimate of each mode; and the \emph{time-weighted mean}~(TWM) is the only table column averaging error over time, computed from the 2-D error of all reported stable estimates and weighted by their time intervals.
Thus the table does not report sample means or 95th-percentile errors, except for the explicitly named TWM summary.
An estimate is declared \emph{converged} when \emph{all} of the following hold simultaneously: $\sigma_H < \SI{2.5}{\metre}$, $\sigma_Z < \SI{8.0}{\metre}$, $n_\text{acc} \ge 15$, the position jump satisfies $\Delta_{xy} < \max(1,\, 2\,\sigma_H)$ and $\Delta_z < \max(2,\, 2\,\sigma_Z)$, and a streak of $\ge 3$ consecutive accepted updates has been observed.
These thresholds relax adaptively once $\sigma_H$ has been non-increasing within a 2\% tolerance for several updates: after five such consecutive updates, $n_\text{acc}$ relaxes to $\ge 10$, and after eight such updates the streak requirement drops to $\ge 2$.
These criteria reflect \emph{filter-internal} consistency and do not bound the absolute positioning error (Section~\ref{sec:overconfidence}).
All algorithmic parameters were set during development and frozen before the field experiments; no parameter was adjusted on the evaluation dataset.

\section{Experimental Setup}\label{sec:setup}

Experiments were conducted on the campus of the University of M\'alaga, Spain, across three days in unstructured, mountainous terrain with dense low vegetation and construction debris.

Three identical Google Nest Wifi routers (model GJ2CQ; IEEE~802.11mc) were deployed at RTK-surveyed positions (\SI{2}{\centi\metre} accuracy), powered by USB battery banks, ranging on the \SI{5}{\giga\hertz} band.
The APs were placed under graduated concealment:
\textbf{AP\_1F} (LoS) on open ground;
\textbf{AP\_AB} (moderate NLoS) hidden under dense vegetation;
\textbf{AP\_AA} (severe NLoS) beneath vegetation and concrete debris.

The u-blox ZED-F9P supported dual-band (L1/L2) GNSS with NTRIP-based RTK corrections via the RAP network.
However, a regional NTRIP outage during the experiments forced standalone operation throughout.
The logged HDOP-derived GNSS quality proxy remained between \num{0.77} and \num{1.02} across \num{14808} fixes, indicating favourable satellite geometry rather than calibrated metre-level accuracy.
Actual horizontal accuracy was approximately \SIrange{1.5}{3}{\metre}, typical for autonomous dual-frequency GNSS~\cite{ublox_f9p}.

\begin{figure}[t]
  \centering
  \includegraphics[width=\columnwidth]{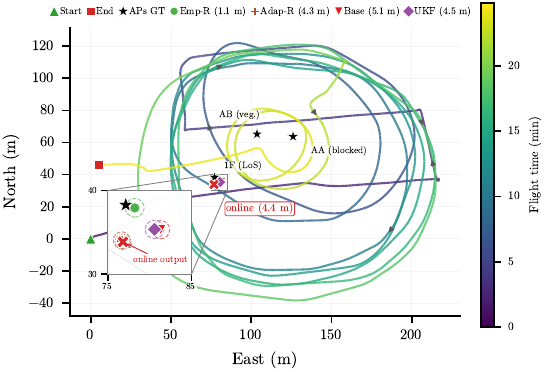}
  \caption{Drone trajectory for the \SI{23}{\minute} flight, with surveyed AP locations and final AP\_1F estimates from offline replay and online Android output; inset: AP\_1F neighbourhood with $\sigma_H$ circles.}
  \label{fig:trajectory}
\end{figure}

The drone (DJI Matrice~350~RTK) flew at \SIrange{35}{55}{\metre} AGL.
The most complete flight (\SI{23}{\minute}, \num{455} FTM measurements to AP\_1F and \num{31} to AP\_AB) was manually teleoperated with circular sweeps to maximise geometric diversity; nearby high-tension power line towers precluded autonomous waypoint missions at the required altitudes, necessitating an experienced pilot.
Figure~\ref{fig:trajectory} shows the drone trajectory in the local ENU frame, with surveyed AP positions and final AP\_1F estimates from the offline replay and online Android output.
The inset zooms into the AP\_1F vicinity, showing the estimates with $\sigma_H$ uncertainty circles.
AP\_AB did not converge due to near-collinear geometry ($\kappa=28.5$) despite \num{31} NLoS detections; AP\_AA produced zero FTM responses (complete signal blockage).

\section{Results}\label{sec:results}

In this section, $\sigma_H$ and $\sigma_Z$ denote filter-internal covariance summaries, not empirical error bounds; absolute errors are computed against the surveyed AP positions.

\subsection{LoS Access Point (AP\_1F)}

Table~\ref{tab:results_1f} summarises the six filter configurations on the longest flight (\num{455} RTT measurements, median range \SI{74}{\metre}).
These values are obtained by offline replay of the recorded RTT and GNSS rosbag, so that all six modes are evaluated on the same measurement sequence.
The raw/derived experimental data and analysis scripts used for this reproducibility package are available on Zenodo at \url{https://doi.org/10.5281/zenodo.20332567}.
The $\varepsilon$ and $\sigma$ columns are final-sample values, while TWM is a time-weighted trajectory summary.
All modes satisfy the filter-internal convergence criteria.
\emph{Empirical-R} achieves the lowest final-sample horizontal error (\SI{1.1}{\metre}; time-weighted mean: \SI{4.7}{\metre}); \textsc{Adaptive-R} reports the smallest internal horizontal uncertainty, with $\sigma_H=\SI{1.0}{\metre}$.
Vertical errors are larger (\SIrange{7.5}{13.8}{\metre}) due to limited altitude diversity.

\emph{Online Android result.}
During the same flight, the Android app ran the real-time \textsc{Baseline} pipeline (MAD gate, IEKF3D, ground-height prior, and median output smoothing) with the \SI{1.2}{\second} GNSS/RTT association window.
The AP\_1F ROS\,2 estimate topic contains \num{270} stable online position estimates; the final message reports \num{283} accepted updates, \SI{4.4}{\metre} horizontal error, \SI{5.5}{\metre} 3-D error, and $\sigma_H=\SI{0.93}{\metre}$.
Across the stable online stream, the median and 95th-percentile errors are \SI{4.2}{\metre}/\SI{5.5}{\metre} in 2-D and \SI{6.1}{\metre}/\SI{8.4}{\metre} in 3-D.
This online result is reported separately from Table~\ref{tab:results_1f} because the table is an offline six-mode replay ablation on common logged inputs, whereas the online stream contains one stability-gated \textsc{Baseline} run with the application's runtime publication logic.

\textbf{Measurement attrition.}
Of \num{455} measurements, \num{22} are consumed by the bootstrap; the remaining $433$ enter the filter.
\textsc{Empirical-R} gates \num{0} (inflated $R_k$ prevents innovation outliers) and accepts all \num{433}, while \textsc{Baseline} rejects \num{94}.
This confirms that model-informed noise scaling preserves measurement diversity, critical for multilateration.
Table~\ref{tab:results_1f} should be interpreted primarily as a final-state comparison; only TWM summarises error over time.

\begin{table}[t]
  \centering
  \caption{Filter comparison on AP\_1F (LoS), \num{455} RTT measurements.  The replay does not add a separate GNSS covariance term to the mode-specific measurement variance.  $n_\text{gate}$: post-bootstrap rejections; $n_\text{acc}$: accepted updates ($n_\text{gate} + n_\text{acc} = 433$).  The $\sigma$ columns are internal covariance standard deviations, not ground-truth errors; $\varepsilon_\text{3D}$ includes the vertical residual.  TWM: time-weighted mean 2-D error computed from the smoothed published estimates.}
  \label{tab:results_1f}
  \setlength{\tabcolsep}{2.5pt}
  \begin{tabular}{@{}lrrrrrrr@{}}
    \toprule
    Mode & $n_\text{gate}$ & $n_\text{acc}$ & $\sigma_H$ [\si{\metre}] & $\sigma_Z$ [\si{\metre}] & $\varepsilon_\text{2D}$ [\si{\metre}] & TWM [\si{\metre}] & $\varepsilon_\text{3D}$ [\si{\metre}]\\
    \midrule
    \textsc{Baseline}    & 94  & 339 & 1.1 & 1.1 &  5.1 & \textbf{4.5} & 13.6 \\
    \textsc{NLoS}        & 78  & 355 & 1.1 & 1.4 &  5.0 & 4.7 & 13.8 \\
    \textsc{Empirical-R} &  0  & 433 & 1.1 & 1.1 & \textbf{1.1} & 4.7 &  \textbf{7.5} \\
    \textsc{UKF}         & 84  & 349 & 1.1 & 1.4 &  4.5 & 5.3 & 13.6 \\
    \textsc{EMA}         & 94  & 339 & 1.1 & 1.1 &  5.0 & 4.6 & 13.8 \\
    \textsc{Adaptive-R}  & 16  & 417 & \textbf{1.0} & 1.1 &  4.3 & 4.9 &  9.6 \\
    \bottomrule
  \end{tabular}
\end{table}

\textbf{Temporal error profile.}
During the first \SI{\sim 17}{\minute} of the post-bootstrap trace, all modes hover around \SIrange{4}{6}{\metre} error; the offline \textsc{Empirical-R} replay has median and 95th-percentile horizontal errors of \SI{4.8}{\metre} and \SI{6.2}{\metre}, respectively, even though its final sample reaches \SI{1.1}{\metre}.
This final improvement occurs only after a close approach: the closest AP\_1F-associated RTT sample is about \SI{11}{\metre} horizontally from the AP roughly \SI{20}{\second} before the final estimate, and the error drops from about \SI{4.5}{\metre} to \SI{1.1}{\metre} over the final \SI{\sim 15}{\second}.
{\scshape Empirical-R} exploits this late geometry because its inflated~$R_k$ preserves more measurements and keeps the covariance amenable to correction; the other modes resist close-pass updates due to tighter, biased covariance.

\begin{figure}[h!]
  \centering
  \includegraphics[width=\columnwidth]{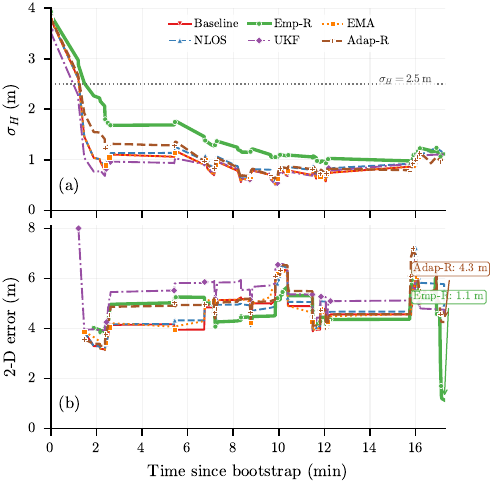}
  \caption{Post-bootstrap offline replay convergence on AP\_1F across six noise configurations: (a)~internal $\sigma_H$ vs.\ time, (b)~ground-truth 2-D error vs.\ time.}
  \label{fig:convergence}
\end{figure}

Figure~\ref{fig:convergence} shows the post-bootstrap offline replay convergence behaviour across all six modes using the AP\_1F measurements from the \SI{23}{\minute} flight.
Panel~(a) plots horizontal uncertainty $\sigma_H$ vs.\ time since bootstrap, with the dashed line marking the $\sigma_H = \SI{2.5}{\metre}$ convergence threshold; this panel reflects internal covariance contraction, not measured accuracy.
Panel~(b) shows 2-D position error vs.\ time; annotated labels highlight the final errors of the two best modes (\textsc{Empirical-R} and \textsc{Adaptive-R}).

\subsection{NLoS Access Points}

AP\_AB (vegetation only) received only \num{31} measurements in \SI{21}{\second}, and all modes failed to converge.
The failure is not due to NLoS alone: sparse detections, signal attenuation, and near-collinear drone positions (condition number $\kappa = 28.5$, vs.\ $2.2$ for AP\_1F) together caused Gauss-Newton divergence despite adequate spatial spread (\SI{28}{\metre} bounding-box diagonal).
A Trust Region Reflective solver (\texttt{scipy.optimize.least\_squares} with Cauchy loss), initialised from the drone centroid and using the same ground-height prior as the online filter, recovers a coarse solution with approximately \SI{7}{\metre} horizontal error.
Thus the vegetation-only data contains useful information, but the online bootstrap lacks the damping and robustness needed under this geometry.
AP\_AA (vegetation plus concrete debris) produced zero FTM responses.

\section{Discussion}\label{sec:discussion}

\subsection{Filter Overconfidence}\label{sec:overconfidence}

Across all modes, $\sigma_H$ is about 4-5$\times$ smaller than $\varepsilon_\text{2D}$, and $\sigma_Z$ is about 6-11$\times$ smaller than vertical error.
The root cause is imperfectly modelled uncertainty in the GNSS-referenced initiator poses: Table~\ref{tab:results_1f} does not include a separate GNSS covariance term, standalone GNSS contains slowly varying correlated errors, and the logged proxy $a_i$ is HDOP-derived rather than a calibrated ENU covariance.
Thus $\sigma_H < \SI{2.5}{\metre}$ indicates internal filter consistency, not a guaranteed \SI{2.5}{\metre} absolute error bound.
Mitigations include RTK-GNSS, propagating calibrated GNSS covariance, estimating a trajectory-level GNSS bias, or inflating $\mathbf{Q}$.

The same distinction explains why the online Android result and the offline replay rows are not numerically identical.
The online run uses the smartphone's configured \textsc{Baseline} mode, runtime GNSS quality proxy, gates, and publication logic; the replay reprocesses the same log under six noise models and omits a separate GNSS covariance term for comparability.
These choices change the accepted-update sequence, Kalman gains, smoothing history, and covariance, so the reported $\sigma_H$ values are internal confidence summaries rather than empirical accuracy estimates from independent trials.

The \SI{1.1}{\metre} \textsc{Empirical-R} final error should likewise be read as a favourable close-pass result rather than a repeatability guarantee.
Slowly varying GNSS biases preserve relative trajectory geometry, but the TWM remains \SI{4.7}{\metre} and the final correction is dominated by late close-approach geometry.
This close-pass pattern is operationally realistic in SAR: the drone can first detect a target from long range, obtain a coarse estimate, and then fly towards the approximate location for a close orbit that refines the position.

\subsection{Bootstrap Geometry and Flight Planning}

The AP\_AB failure demonstrates that bounding-box spread alone is insufficient to guarantee solver convergence: the bootstrap should also check the geometry's condition number ($\kappa < 10$ would have correctly rejected AP\_AB while accepting AP\_1F).
The AP\_1F close-pass effect shows that final accuracy can be dominated by one favourable geometry event, while manual piloting limits repeatability through variations in antenna position, attitude, speed, and turn geometry.
Autonomous lawnmower, grid/mesh, helical, or concentric-circle trajectories should therefore plan closest approach, bearing diversity, and altitude variation jointly.

\subsection{Towards Networked Multi-Agent Search}\label{sec:multi_osa}

The single-agent pipeline is designed as the building block of a Networked OSA architecture, where each OSA is still a single self-contained search agent but multiple OSAs share FTM observations and GNSS-referenced positions via ROS\,2 for centralised fusion.
Preliminary experiments with a UAV and a human operator across five sessions processed up to \num{1802} RTT measurements from the ground agent and produced \num{992} fused position estimates, demonstrating software-level interoperability with heterogeneous agents.
However, incorporating measurements of differing quality without proper weighting can degrade the fused estimate; a rigorous quantitative evaluation is left for future work.


\section{Conclusions and Future Work}\label{sec:conclusions}

This paper demonstrated that a UAV equipped with a commodity smartphone and an external GNSS receiver can localise ground-based WiFi FTM responders using GNSS-referenced moving-baseline multilateration.
On AP\_1F, the logged Android \textsc{Baseline} pipeline achieved \SI{4.4}{\metre} final horizontal error and \SI{5.5}{\metre} 3-D error, while offline replay with the empirical distance- and RSSI-dependent noise model achieved \SI{4.7}{\metre} time-weighted mean horizontal error and a best-case \SI{1.1}{\metre} final horizontal error after a late close approach.
This final value is a favourable-case outcome rather than a guaranteed system bound, because the GNSS-referenced initiator poses carried metre-level uncertainty whose correlated component was not captured by the FTM-only covariance model used in the comparison.

The experiments also highlight the importance of flight geometry.
Useful horizontal bearing diversity was achievable with circular sweeps and close passes, but limited altitude variation left the AP height weakly observable and produced vertical errors of several metres.
Future search trajectories should therefore optimise both area coverage and multilateration conditioning, for example through lawnmower scans, grid-based trajectories, helical approaches, or concentric circles with planned altitude variation.

NLoS operation remains the main unresolved challenge.
Vegetation-only concealment produced intermittent FTM detections, and offline robust least-squares with weak height regularisation recovered a coarse solution with approximately \SI{7}{\metre} horizontal error, indicating that the measurements still contained useful information.
However, the current online Gauss-Newton bootstrap failed under near-degenerate geometry, while vegetation combined with concrete debris completely blocked FTM responses.
Therefore, future implementations should replace the current bootstrap with a damped robust solver, include explicit geometry checks before declaring convergence, and incorporate NLoS-aware measurement weighting.

The single-agent pipeline is intended as a building block for Networked OSA systems.
Follow-up work using JEMERG~2026 data will evaluate network-level fusion with multiple UAVs, calibrated inter-agent weighting, and integration with heterogeneous DJI platforms through WildBridge-style adapters~\cite{rolland2025wildbridge} and Swarm-Steward supervision~\cite{jarabo2026swarmsteward}.

\section*{Acknowledgements}
This work was partially supported by the Spanish Ministerio de Ciencia, Innovaci\'on y Universidades (project PID2021-122944OB-I00), University of M\'alaga (UMA), Innovation Fund Denmark (DIREC U07, PERSIST), and the Independent Research Fund Denmark (Grant 10.46540/4264-00105B, NAMUR).
The authors also thank the C\'atedra de Seguridad, Emergencias y Cat\'astrofes at the UMA for organising the JEMERG field exercises~\cite{jemerg2026}, which provided the operational context for this work.

\balance


\begin{thebibliography}{99}



\bibitem{lyu2023uav}
M.~Lyu, Y.~Zhao, C.~Huang, and H.~Huang,
``Unmanned aerial vehicles for search and rescue: A survey,''
\emph{Remote Sensing}, vol.~15, no.~13, p.~3266, 2023.

\bibitem{queralta2020multi}
J.\,P.~Queralta, J.~Taipalmaa, B.\,C.~Pullinen, V.\,K.~Sarker, T.\,N.~Gia, H.~Tenhunen, M.~Gabbouj, J.~Raitoharju, and T.~Westerlund,
``Collaborative multi-robot search and rescue: Planning, coordination, perception, and active vision,''
\emph{IEEE Access}, vol.~8, pp.~191617--191643, 2020.

\bibitem{jemerg2026}
Universidad de M\'alaga, C\'atedra de Seguridad, Emergencias y Cat\'astrofes,
``XX Jornadas Internacionales de la Universidad de M\'alaga sobre Seguridad, Emergencias y Cat\'astrofes: Protocolos y Capacidades para la Defensa del Patrimonio Cultural y Natural ante Emergencias,''
M\'alaga, Spain, May~7--8, 2026.
[Online]. Available: \url{https://www.jornadascatastrofes.com/}

\bibitem{pagliari2024wi}
E.~Pagliari, L.~Davoli, and G.~Ferrari,
``Wi-Fi-based real-time UAV localization: A comparative analysis between RSSI-based and FTM-based approaches,''
\emph{IEEE Trans. Aerosp. Electron. Syst.}, vol.~60, no.~6, pp.~8757--8778, 2024.

\bibitem{alarifi2016uwb}
A.~Alarifi \emph{et al.},
``Ultra wideband indoor positioning technologies: Analysis and recent advances,''
\emph{Sensors}, vol.~16, no.~5, p.~707, 2016.

\bibitem{rykala2020research}
{\L}.~Ryka{\l}a, A.~Typiak, and R.~Typiak,
``Research on developing an outdoor location system based on the ultra-wideband technology,''
\emph{Sensors}, vol.~20, no.~21, p.~6171, 2020.

\bibitem{queralta2020uwb}
J.\,P.~Queralta, C.~Almansa, F.~Schiano, D.~Floreano, and T.~Westerlund,
``UWB-based system for UAV localization in GNSS-denied environments: Characterization and dataset,''
in \emph{Proc.\ IEEE/RSJ IROS}, 2020, pp.~4521--4528.

\bibitem{schroth2024emergency}
C.\,A.~Schroth, C.~Eckrich, I.~Kakouche, S.~Fabian, O.~von~Stryk, A.\,M.~Zoubir, and M.~Muma,
``Emergency response person localization and vital sign estimation using a semi-autonomous robot mounted SFCW radar,''
\emph{IEEE Trans. Biomed. Eng.}, vol.~71, no.~6, pp.~1756--1769, 2024.

\bibitem{ftm_standard}
IEEE Std 802.11-2020,
``Part 11: Wireless LAN MAC and PHY specifications,'' IEEE, 2021.

\bibitem{alvarez2021wifi}
C.~S.~\'Alvarez-Merino, H.\,Q.~Luo-Chen, E.\,J.~Khatib, and R.~Barco,
``WiFi FTM, UWB and cellular-based radio fusion for indoor positioning,''
\emph{Sensors}, vol.~21, no.~21, p.~7020, 2021.

\bibitem{android_rtt}
Android Open Source Project,
``Wi-Fi RTT (Round Trip Time),''
[Online]. Available: \url{https://source.android.com/docs/core/connect/wifi-rtt}

\bibitem{yu2020precise}
Y.~Yu, R.~Chen, L.~Chen, S.~Xu, W.~Li, Y.~Wu, and H.~Zhou,
``Precise 3-D indoor localization based on Wi-Fi FTM and built-in sensors,''
\emph{IEEE Internet Things J.}, vol.~7, no.~12, pp.~11753--11765, 2020.

\bibitem{ibrahim2018}
M.~Ibrahim, H.~Liu, M.~Jawahar, V.~Nguyen, M.~Gruteser, R.~Howard, B.~Yu, and F.~Bai,
``Verification: Accuracy evaluation of WiFi fine time measurements on an open platform,''
in \emph{Proc.\ ACM MobiCom}, 2018, pp.~417--427.

\bibitem{ma2020wifi}
C.~Ma, B.~Wu, S.~Poslad, and D.\,R.~Selviah,
``Wi-Fi RTT ranging performance characterization and positioning system design,''
\emph{IEEE Trans. Mobile Comput.}, vol.~21, no.~2, pp.~740--756, 2022.

\bibitem{martin2020passive}
I.~Martin-Escalona and E.~Zola,
``Passive round-trip-time positioning in dense IEEE 802.11 networks,''
\emph{Electronics}, vol.~9, no.~8, p.~1193, 2020.

\bibitem{yu2021seamless}
Y.~Yu, R.~Chen, L.~Chen, W.~Li, Y.~Wu, and H.~Zhou,
``A robust seamless localization framework based on Wi-Fi FTM/GNSS and built-in sensors,''
\emph{IEEE Commun. Lett.}, vol.~25, no.~7, pp.~2226--2230, 2021.

\bibitem{horn2024outdoor}
B.\,K.\,P.~Horn,
``Round-trip time ranging to Wi-Fi access points beats GNSS localization,''
\emph{Appl. Sci.}, vol.~14, no.~17, p.~7805, 2024.

\bibitem{lu2024dnn}
B.~Lu, M.~Wang, W.~Wen, and Y.~Zhang,
``Improving FTM ranging accuracy based on DNN for UAV localization,''
\emph{IEEE Internet Things J.}, vol.~11, no.~12, pp.~21287--21298, 2024.

\bibitem{li2026coloc}
L.~Li, M.~Wang, Y.~Wang, F.~Gu, L.~Chen, R.~Chen, M.~Liu, and S.~Jin,
``Pedestrian and router colocalization framework using distributed IMU-based VDR and Wi-Fi RTT,''
\emph{IEEE Internet of Things Journal}, early access, 2026,
doi:~10.1109/JIOT.2026.3656409.

\bibitem{shaikhanov2022falcon}
Z.~Shaikhanov, A.~Boubrima, and E.\,W.~Knightly,
``FALCON: A networked drone system for sensing, localizing, and approaching RF targets,''
\emph{IEEE Internet Things J.}, vol.~9, no.~12, pp.~9843--9857, 2022.

\bibitem{bravo2022realistic}
J.~Bravo-Arrabal, P.~Zambrana, J.\,J.~Fern\'andez-Lozano, J.\,A.~G\'omez-Ruiz, J.~Ser\'on-Barba, and A.~Garc\'\i a-Cerezo,
``Realistic deployment of hybrid wireless sensor networks based on ZigBee and LoRa for search and rescue applications,''
\emph{IEEE Access}, vol.~10, pp.~64618--64637, 2022.

\bibitem{alvarez2022victim}
C.~S.~\'Alvarez-Merino, E.\,J.~Khatib, H.\,Q.~Luo-Chen, and R.~Barco,
``Victim detection and localization in emergencies,''
\emph{Sensors}, vol.~22, no.~21, p.~8433, 2022.

\bibitem{khatib2023designing}
E.\,J.~Khatib, C.\,S.~\'Alvarez-Merino, H.\,Q.~Luo-Chen, and R.~Barco,
``Designing a 6G testbed for location: Use cases, challenges, enablers and requirements,''
\emph{IEEE Access}, vol.~11, pp.~10053--10091, 2023.

\bibitem{bravo2025real}
J.~Bravo-Arrabal, C.\,S.~\'Alvarez-Merino, M.~Toscano-Moreno, J.~Ser\'on, J.\,J.~Fern\'andez-Lozano, J.\,A.~G\'omez-Ruiz, E.\,J.~Khatib, R.~Barco, and A.~Garc\'\i a-Cerezo,
``Real-time FTM-based victim positioning system using heterogeneous robots in remote and outdoor scenarios,''
\emph{IEEE Access}, vol.~13, pp.~86949--86967, 2025.

\bibitem{bravo2026osa}
J.~Bravo-Arrabal, C.\,S.~\'Alvarez-Merino, M.~Toscano-Moreno, J.~Ser\'on-Barba, J.\,J.~Fern\'andez-Lozano, J.\,A.~G\'omez-Ruiz, E.\,J.~Khatib, R.~Barco, and A.~Garc\'\i a-Cerezo,
``Real-time victim positioning with one search agent using Wi-Fi fine time measurement,''
\emph{IEEE Systems Journal}, under review, 2026.

\bibitem{ramos2024ur2a}
M.~C\'ordoba-Ramos, J.~Bravo-Arrabal, J.\,J.~Fern\'andez-Lozano, A.~Mandow, and A.~Garc\'\i a-Cerezo,
``UR2A: comunicaci\'on bidireccional Android-ROS\,2 para arquitecturas edge-cloud en sistemas rob\'oticos conectados,''
\emph{Jornadas de Autom\'atica}, no.~45, 2024,
doi:~10.17979/ja-cea.2024.45.10896.

\bibitem{huber2009}
P.\,J.~Huber and E.\,M.~Ronchetti,
\emph{Robust Statistics}, 2nd~ed.
Hoboken, NJ, USA: Wiley, 2009.

\bibitem{iglewicz1993}
B.~Iglewicz and D.\,C.~Hoaglin,
``How to detect and handle outliers,''
in \emph{The ASQC Basic References in Quality Control: Statistical Techniques}, vol.~16.\hspace{1mm}ASQC/Quality Press, 1993.

\bibitem{mohamed1999}
A.\,H.~Mohamed and K.\,P.~Schwarz,
``Adaptive Kalman filtering for INS/GPS,''
\emph{Journal of Geodesy}, vol.~73, no.~4, pp.~193--203, 1999,
doi:~10.1007/s001900050236.

\bibitem{ublox_f9p}
u-blox,
``ZED-F9P-01B u-blox F9 high precision GNSS module: Data sheet,''
document no.~UBX-17051259, Rev.~R09, Mar.~2023.
[Online]. Available: \url{https://www.u-blox.com/en/product/zed-f9p-module}

\bibitem{rolland2025wildbridge}
E.~Rolland \emph{et al.},
``WildBridge: Ground station interface for lightweight multi-drone control and telemetry on DJI platforms,''
in \emph{Proc.\ 13th Int. Conf. Robot Intelligence Technology and Applications (RiTA 2025)},
London, United Kingdom: Springer, Dec.~2025, in press.

\bibitem{jarabo2026swarmsteward}
A.~Jarabo-Pe\~nas, J.~Bravo-Arrabal, E.\,G.\,A.~Rolland, and A.\,L.~Christensen,
``Swarm-Steward: Scalable and reliable natural-language coordination of autonomous aerial and ground robots,''
accepted for publication in \emph{Proc.\ Int. Conf. Unmanned Aircraft Systems (ICUAS)}, 2026.

\bibitem{liu2024tightly}
T.~Liu, B.~Li, G.~Chen, L.~Yang, J.~Qiao, and W.~Chen,
``Tightly coupled integration of GNSS/UWB/VIO for reliable and seamless positioning,''
\emph{IEEE Trans. Intell. Transp. Syst.}, vol.~25, no.~2, pp.~2116--2128, 2024.

\end{thebibliography}
\end{document}